\documentclass[preprint2]{aastex}

\slugcomment{To appear in ApJ Letters}

\shorttitle{Faint Herbig-Haro Jet}
\shortauthors{Riaz \& Whelan}

\begin{document}

\title{HH 1158: The lowest luminosity externally irradiated Herbig-Haro jet}

\author{B. Riaz}
\affil{Max-Planck-Institut f\"{u}r Extraterrestrische Physik, Giessenbachstrasse 1, 85748 Garching, Germany}

\and

\author{E. T. Whelan\altaffilmark{1}}
\affil{Institute f\"{u}r Astronomie und Astrophysik, Eberhard Karls University Tuebingen, Sand 1, D-72076 T\"{u}bingen, Germany}

\altaffiltext{1}{National University of Ireland, Maynooth, Ireland}

\begin{abstract}

We have identified a new externally irradiated Herbig-Haro (HH) jet, HH 1158, within $\sim$2 pc of the massive OB type stars in the $\sigma$ Orionis cluster. At an $L_{bol}$ $\sim$ 0.1 $L_{\sun}$, HH 1158 is the lowest luminosity irradiated HH jet identified to date in any cluster. Results from the analysis of high-resolution optical spectra indicate asymmetries in the brightness, morphology, electron density, velocity, and the mass outflow rates for the blue and red-shifted lobes. We constrain the position angle of the HH 1158 jet at 102$^{\circ}$$\pm$5$^{\circ}$. The mass outflow rate and the mean accretion rate for HH 1158 using multiple diagnostics are estimated to be (5.2$\pm$2.6) x10$^{-10}$ M$_{\sun}$ yr$^{-1}$ and (3.0$\pm$1.0) x 10$^{-10}$ M$_{\sun}$ yr$^{-1}$, respectively. The properties for HH 1158 are notably similar to the externally irradiated HH 444 -- HH 447 jets previously identified in $\sigma$ Orionis. In particular, the morphology is such that the weaker jet beam is tilted towards the massive stars, indicating a higher extent of photo-evaporation. The high value for the H$\alpha$/[S~{\sc ii}] ratio is also consistent with the ratios measured in other irradiated jets, including HH 444 -- HH 447.  The presence of an extended collimated jet that is bipolar and the evidence of shocked emission knots make HH 1158 the first unique case of irradiated HH jets at the very low-luminosity end, and provides an opportunity to learn the physical properties of very faint HH jet sources. 

\end{abstract}


\section{Introduction} \label{intro}

Herbig-Haro (HH) objects are collisionally excited nebulae produced by highly collimated outflows ejected from young stellar objects (YSOs). Most HH jet sources are embedded within opaque cloud cores that obscure the immediate vicinity of the outflow sources. A case in exception are {\it externally irradiated HH jets} identified in H II regions or located near massive OB type stars. The external ultraviolet (UV) radiation field ionizes and photoablates the bulk of the gas surrounding the YSOs, and contributes to the excitation and visibility of the outflow (Reipurth et al. 1998; Reipurth \& Bally 2001). 

\begin{figure}
%\center
\includegraphics[scale=.60]{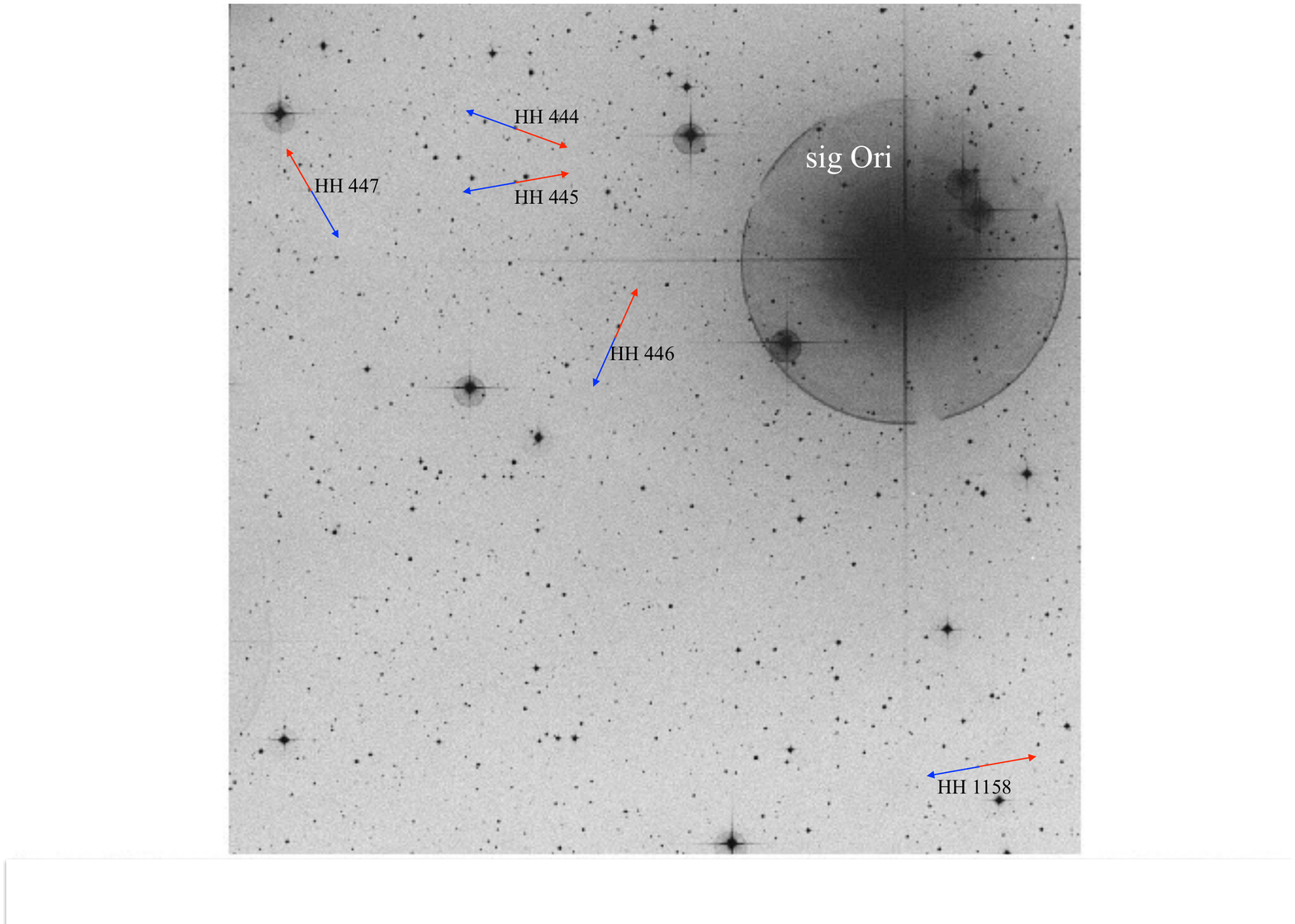}
\caption{The UKIDSS $K_{s}$-band image showing $\sigma$Ori, the HH 444 -- HH 447, and HH 1158 jets. The orientation of the blue and red lobes for the jets are marked as blue and red arrows, respectively. The scale of the image is about 30$\arcmin$x30$\arcmin$. North is up, East is to the left. }
\label{field}
\end{figure}

Among the first such cases of externally irradiated HH jets were HH 444 -- HH 447, identified in the $\sigma$ Orionis ($\sigma$ Ori) cluster (Reipurth et al. 1998). The $\sigma$ Ori cluster belongs to the Orion OB 1b association. It is centered around the massive O9.5-type multiple star system of the same name, and has been suggested as an H II region (e.g., Walter et al. 2008). The HH 444 -- HH 447 jets are located within $\sim$2 pc of $\sigma$ Ori (Fig.~\ref{field}), and are driven by optically visible T Tauri stars. Interestingly, all four jets are highly asymmetric, with the beam facing towards $\sigma$ Ori being much weaker and fainter than the counter beam, indicating a higher extent of photo-evaporation due to the ionizing radiation from the massive stars (Reipurth et al. 1998; Andrews et al. 2004). The sources of the irradiated jets are considered to be in an evolutionary state where they have retained tenuous or no envelope material, thereby exposing the disks to photo-erosion. The production of jets seems to have been unaffected. The main utility of such irradiated jets is that the radiation field can improve their visibility both close to and far from the shock region, thus enabling unambiguous determination of the jet physical parameters.

In this paper, we report the identification of a new irradiated jet, HH 1158, in the $\sigma$ Ori cluster. The driving source of HH 1158 is a very low-luminosity ($L_{bol}$ = 0.16$\pm$0.03 $L_{\sun}$) star, named Mayrit 1082188 (M1082188). In Riaz et al. (2015), we presented multi-wavelength optical through sub-millimeter observations for this YSO, including low-resolution optical spectroscopy. M1082188 is classified as a Flat Spectrum object. The Class Flat sources are considered to be at an intermediate stage between Class I and II and have tenuous envelopes compared to Class I objects (Greene et al. 1994). The proximity (within $\sim$2 pc) of M1082188 to the massive $\sigma$ Ori quintuple system (Fig.~\ref{field}) suggests that the ionizing radiation from nearby OB stars has stripped the circumstellar envelope surrounding the driving source of this irradiated jet, making it visible at optical and near-infrared wavelengths. The total (dust+gas) mass for this system as derived from the sub-millimeter fluxes is $\sim$22 $M_{Jup}$. The optical spectrum exhibits prominent emission in the [O~{\sc i}] $\lambda$$\lambda$6300, 6363\,{\AA}, [S~{\sc ii}] $\lambda$$\lambda$6716, 6730\,{\AA}, and the [N~{\sc ii}] $\lambda$ 6583$\AA$ forbidden emission lines (FELs), indicating that it is driving an outflow. Here, we present new results from an analysis of high-resolution spectra for HH 1158 obtained at the Very Large Telescope (VLT), using the UV-Visual Echelle Spectrometer (UVES). Interestingly, HH 1158 shows a similar bipolar asymmetric jet morphology as observed for the HH 444 -- HH 447 jets, and is the lowest luminosity HH jet identified yet in the $\sigma$ Ori or any other cluster. While numerous example of jets from Class II very low-mass stars and brown dwarfs have now been studied (e.g., Whelan et al. 2009), these jets are considered to be analogous to the ``micro-jets'' driven by classical T Tauri stars, and no definite detection of HH objects associated with these jets have been made  prior to this work. Thus this work provides us with a new way in which to study HH jet phenomena at the lowest masses. In Section~\ref{obs}, we give details of the new observations, the results from the spectral analysis are presented in Section~\ref{results}, and the similarities between HH 1158 and HH 444 -- HH 447 jets are discussed in Section \ref{discuss}.

\section{Observations and Data Reduction} \label{obs}

\begin{figure}
\center
\includegraphics[scale=0.6]{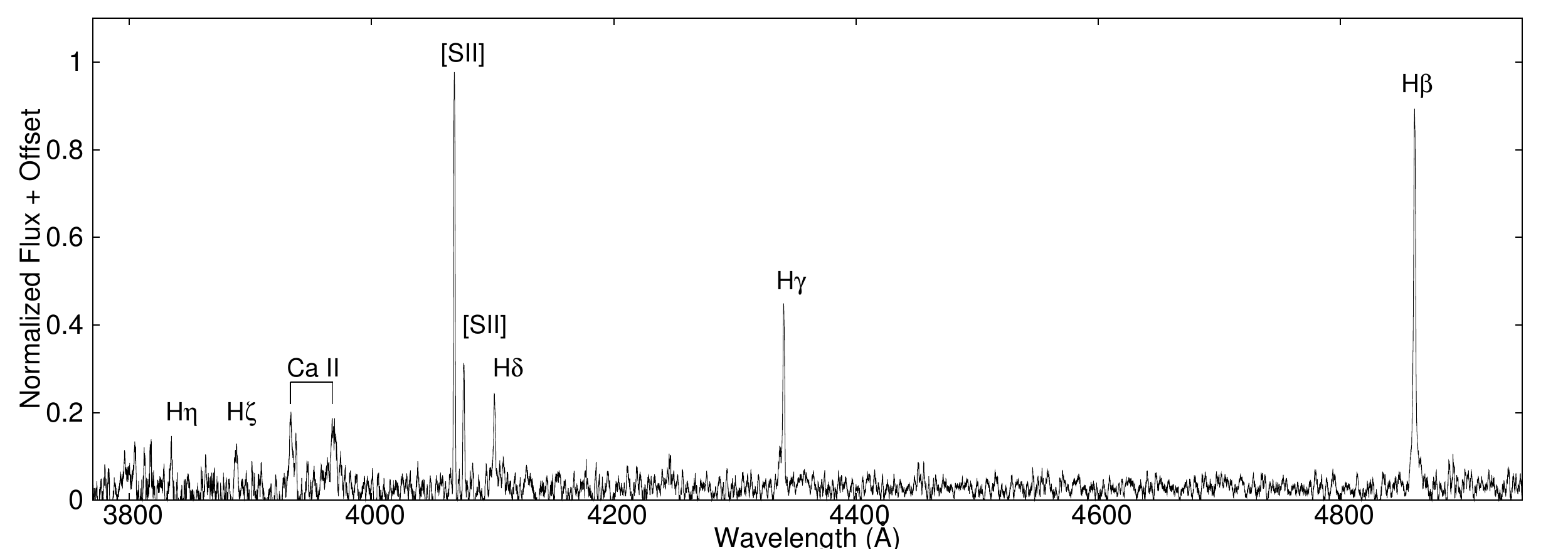}
\includegraphics[scale=0.6]{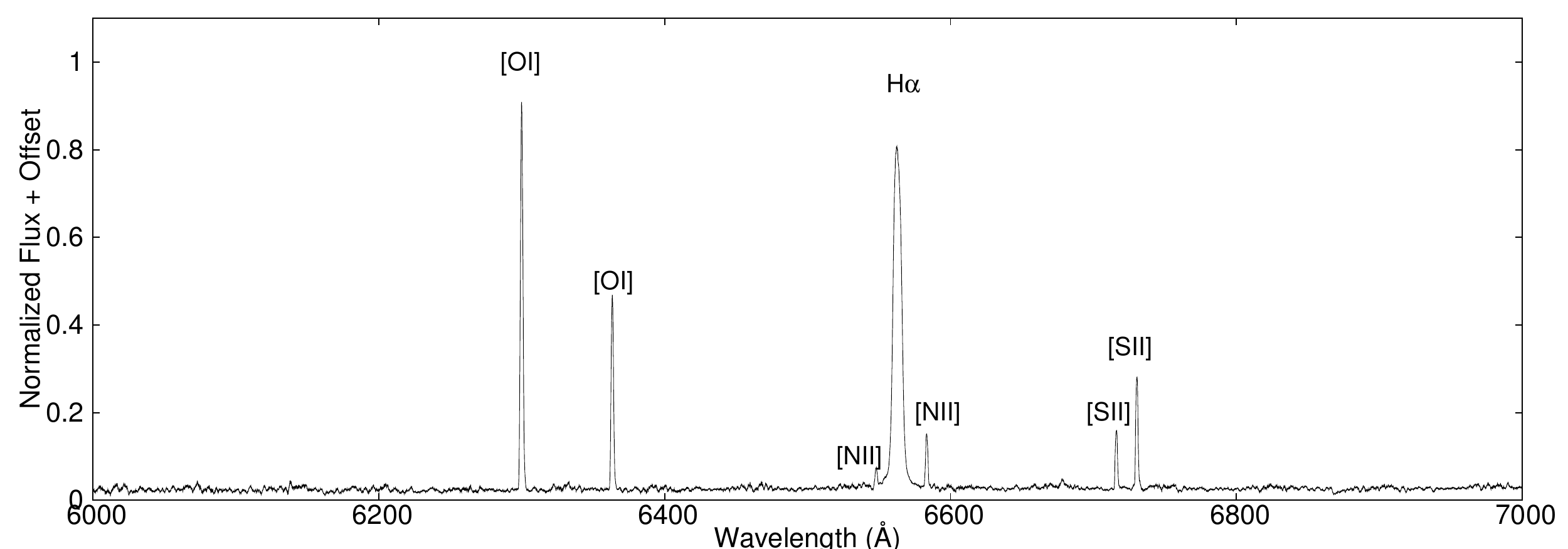}
\includegraphics[scale=0.6]{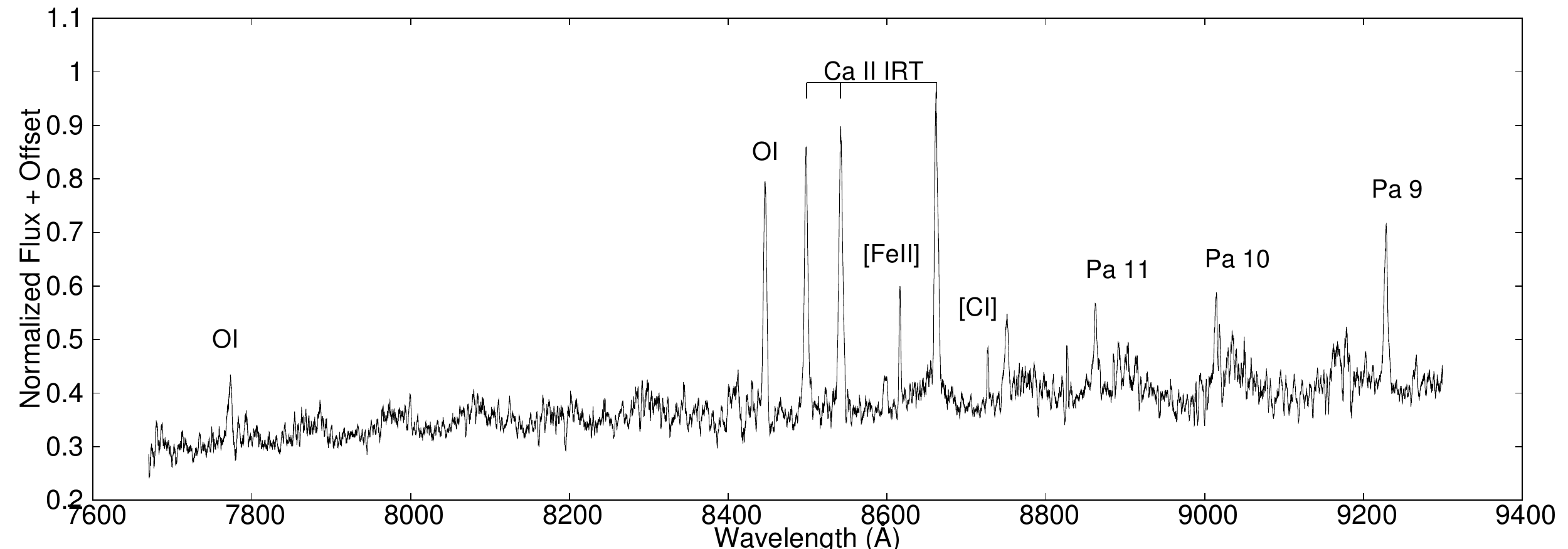}
\caption{UVES spectra for HH 1158 with the prominent accretion- and outflow-associated emission lines marked. }
\label{fullspec}
\end{figure}

The UVES high-resolution spectra for HH 1158 (Program ID: 094.C-0667(A)) were obtained in September, 2014. We used the standard DIC2 (437+760) setting, with cross dispersers of CD\#2 (HER\_5) and CD\#4 (BK7\_5). This setup provided a wavelength coverage from 373 to 946 nm in one exposure. The slit width was set to 1$\arcsec$, resulting in a spectral resolution of $R \sim$ 40,000. We obtained two spectra for the target, one per slit orientation at a position angle (P.A.) of 0$\degr$ and 90$\degr$. The total on-source exposure time was 3200 seconds, split into two exposures. The seeing was recorded to be between 0.5$\arcsec$ and 1$\arcsec$ during the nights when these observations were made. The UVES spectra were reduced using the ESO Reflex pipeline. The combined 0$\degr$ and 90$\degr$ spectra are shown in Fig.~\ref{fullspec}; the fluxes measured in notable accretion and outflow diagnostics are listed in Table~\ref{fluxes}. We estimate a signal-to-noise ratio (SNR) of better than $\sim$20 in the red arm and $\sim$10 in the blue part of the combined spectrum. 

For the position-velocity (PV) diagrams and the spectro-astrometric analysis, we have used the standard IRAF routines for continuum and sky line subtraction and for Gaussian fitting. Spectro-astrometry is a technique by which Gaussian fitting of the spatial profile of a spectrum as a function of wavelength is used to recover spatial information from spatially unresolved emission lines. It has been successfully used to disentangle outflow components to emission lines tracing accretion and to detect brown dwarf outflows (Whelan et al. 2005). We have followed the same process for the spectro-astrometric analysis as outlined in (Whelan \& Garcia 2008).

\section{Results} \label{results}

The optical spectrum for HH 1158 exhibits all of the well-known outflow-associated [O~{\sc i}], [S~{\sc ii}], [N~{\sc ii}], and [Fe~{\sc ii}] FELs (Fig.~\ref{fullspec}). There is also a marginal detection of the [N~{\sc ii}] $\lambda$ 6548\,{\AA} feature clearly resolved from the H$\alpha$ line. This FEL is rarely seen in spectra of very low-mass stars or brown dwarfs, as it requires a higher temperature and density than the [N~{\sc ii}] $\lambda$ 6583\,{\AA} line. HH 1158 has strong H~{\sc i} lines including the Balmer series (H$\eta$, H$\zeta$, H$\delta$, H$\gamma$, H$\beta$, H$\alpha$), and the Paschen series (Pa 9, 10, 11). Also notable is emission in the Ca~{\sc ii} H and K lines and the Ca~{\sc ii} infrared triplet, typically associated with strong accretion. Another unique feature observed is strong emission in the O~{\sc i} $\lambda$$\lambda$7773, 8446\,{\AA} atomic lines, which are suggested to be kinematically associated with a strong outflow, and possibly have an origin in high-velocity winds (e.g., Hillenbrand et al. 2012). 

%The Ca~{\sc ii} H line is blended with the H$\epsilon$ line, and the Pa 9 line is blended with the [Fe~{\sc ii}] line at $\sim$9227\,{\AA}. 

\begin{figure}
%\epsscale{0.08}
%\plottwo{f3a.pdf}{f3b.pdf}
\includegraphics[scale=0.3]{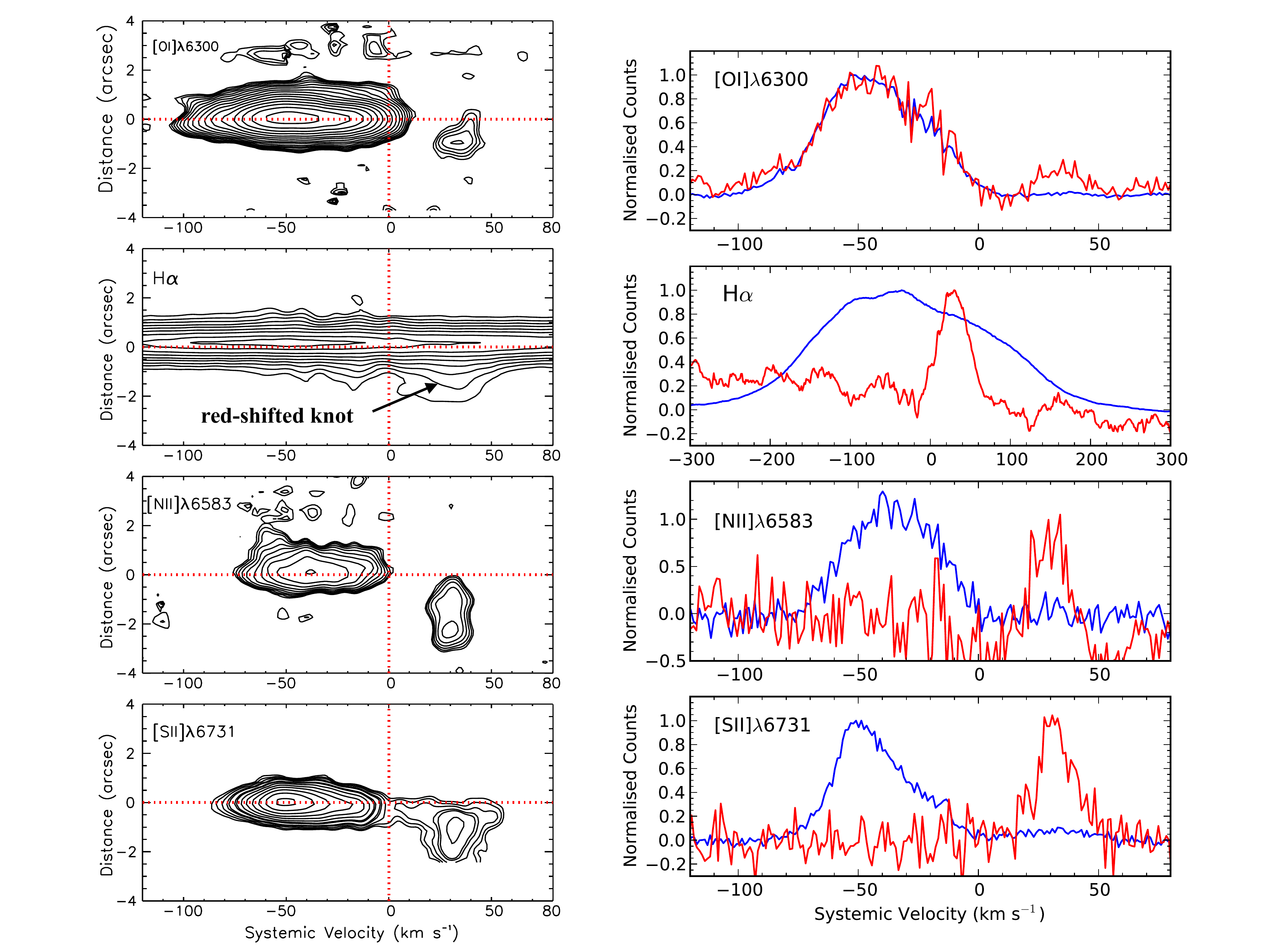} 
\includegraphics[scale=0.4]{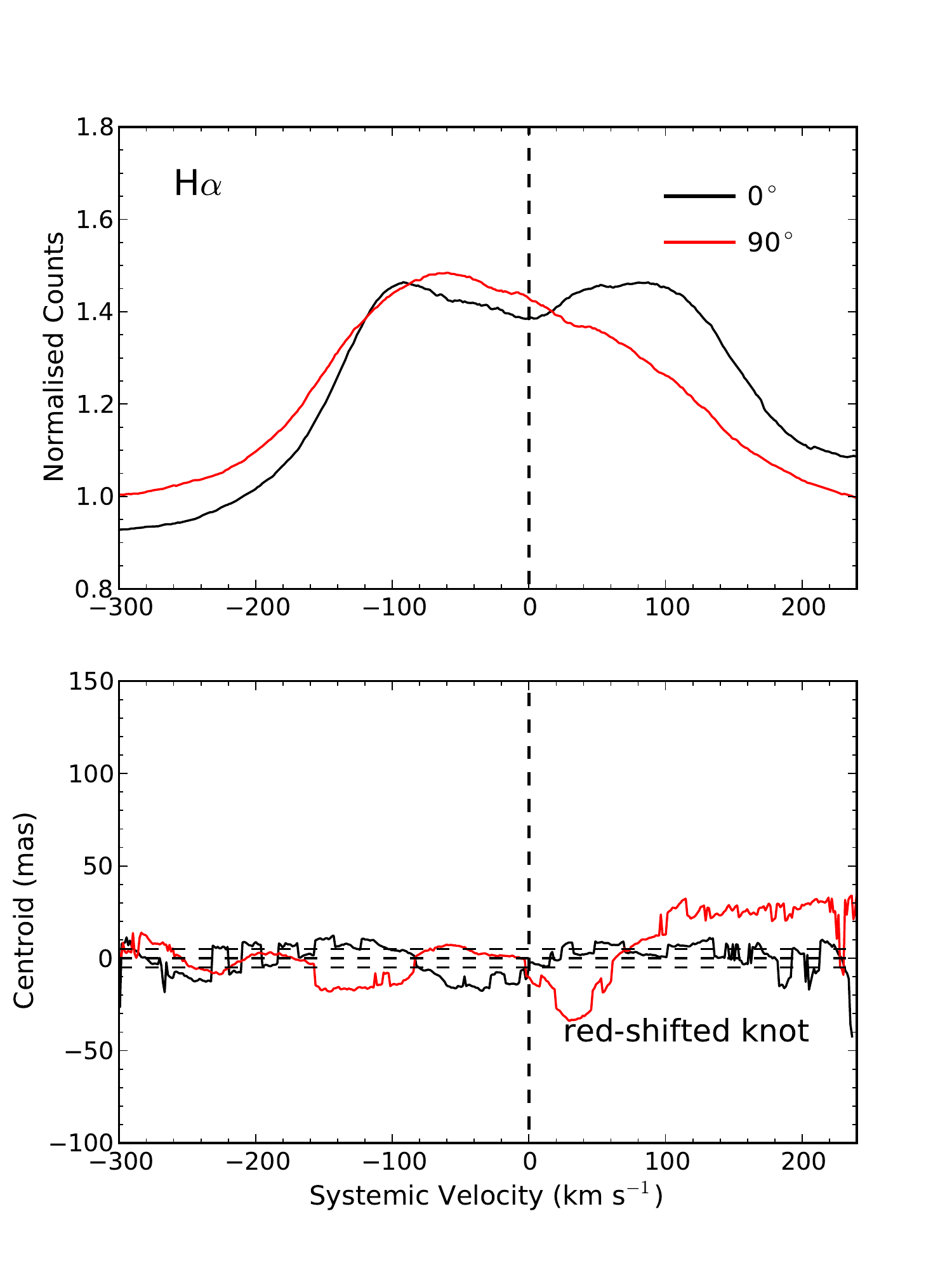}
\caption{{\bf {\it Left} (a.)} The PV plots for HH 1158 alongside the line profiles, constructed from the 90$^{\circ}$ spectra. The line profiles are extracted from a position of 0$\arcsec$ and -1.5$\arcsec$. The extracted lines have been normalized to the line peaks. {\bf {\it Right} (b.)}  Spectro-astrometric analysis of the HH 1158 H$\alpha$ line. The spectra have been binned to increase the SNR and the spectro-astrometric accuracy. The horizontal lines delinates the $\pm$1-$\sigma$ uncertainty. A small offset is seen at $\sim$20 km s$^{-1}$, which is the red-shifted lobe of the jet.  }
\label{pvdiags}
\end{figure}

The position-velocity diagrams of the continuum-subtracted [O~{\sc i}] $\lambda$6300, [N~{\sc ii}] $\lambda$6584, [S~{\sc ii}] $\lambda$6731, and H$\alpha$ lines from the HH 1158 jet in the 90$^{\circ}$ spectrum are shown in Fig.~\ref{pvdiags}a. Similar results are obtained in the [O~{\sc i}] $\lambda$6363, [N~{\sc ii}] $\lambda$6546, [S~{\sc ii}] $\lambda$6716 lines. The velocities shown are systemic; we have considered -30.9 km s$^{-1}$ for the systemic velocity (Riaz et al. 2015). Also shown in Fig.~\ref{pvdiags}a alongside the PV diagrams are the respective line profiles that have been extracted at a position of 0$\arcsec$ (blue) and -1.5$\arcsec$ (red). The jet for HH 1158 is spatially resolved at a P. A. of 90$^{\circ}$; we see spatially extended emission at this P. A., while the emission is not spatially resolved in the 0$^{\circ}$ spectrum. To map the offset of the emission at 0$^{\circ}$, we have applied spectro-astrometry and measured offsets of $\sim$200 mas in the red-shifted jet, and $\sim$50 mas in the blue-shifted jet. This is consistent with the spectrum at 90$^{\circ}$ that shows the red-shifted emission to be positioned much further from the star than the blue-shifted emission. Using the results from the 90$^{\circ}$ and 0$^{\circ}$ spectra, we constrain the PA of the HH 1158 jet at 102$^{\circ}$$\pm$5$^{\circ}$.

HH 1158 exhibits a bipolar asymmetric jet. The blue-shifted lobe is brighter and has a slightly higher velocity than the red-shifted component. The difference in brightness is well illustrated by the contour plots. Note that the line profiles have been normalized to the peak value since the blue emission is much brighter than the red one, and so the profiles provide information on the relative shape and velocity but not the relative brightness of the components. The line flux ratios in the FELs measured from the 90$^{\circ}$ spectrum in the blue and red lobes are approximately [O~{\sc i}]\_blue / [O~{\sc i}]\_red = 67, [N~{\sc ii}]\_blue / [N~{\sc ii}]\_red = 4, and [S~{\sc ii}]\_blue / [S~{\sc ii}]\_red = 8.5, which clearly indicate the relative brightness of the blue-shifted emission. Using the [S~{\sc ii}] line ratios, we derive electron densities, $n_{e}$, of 4000 cm$^{-3}$ and 1000 cm$^{-3}$ in the blue and red lobe, respectively. Therefore, the jet is also asymmetric in density. 

The red-shifted emission is more spatially extended ($\sim$2$\arcsec$--3$\arcsec$), compared to the blue component. The jet is much broader in velocity in [O~{\sc i}] than in the [S~{\sc ii}] line. The [O~{\sc i}]$\lambda$6300 line has a larger critical density and traces the jet closer to the source than the other FELs. The [S~{\sc ii}] emission, in comparison, can trace the jet knots farther out from the driving source and is more extended. The [O~{\sc i}]\_blue line probably has a contribution from a low velocity wind close to the source, which is not seen in the red due to partial obscuration by the disk. This could explain why the blue component is broader than the red lobe. This is also perhaps why we do not see the first knot of red-shifted emission until around 1$\arcsec$ from the driving source. 

The H$\alpha$ line shows a broad profile mainly arising from the stellar centroid at 0$\arcsec$ position. The red-shifted jet emission seen in the FELs is also detected in the H$\alpha$ line at $\sim$20-30 km s$^{-1}$, at a position of -1.5$\arcsec$. The H$\alpha$ line is a better probe of diffuse jet regions and the irradiated portions, therefore some contribution from the jet emission can be expected. We have conducted a spectro-astrometric analysis in order to investigate the presence of blue-shifted jet emission in the wing of the H$\alpha$ line. Figure~\ref{pvdiags}b shows a small offset at the position of the red-shifted H$\alpha$ knot. However, no blue-shifted jet emission is seen, and the bulk of the H$\alpha$ emission is tracing accretion. This is also notable from the broad H$\alpha$ profile, with a width at 10\% of the line peak of 610$\pm$30 km s$^{-1}$, indicating intense accretion.

\begin{figure}
\includegraphics[scale=0.5]{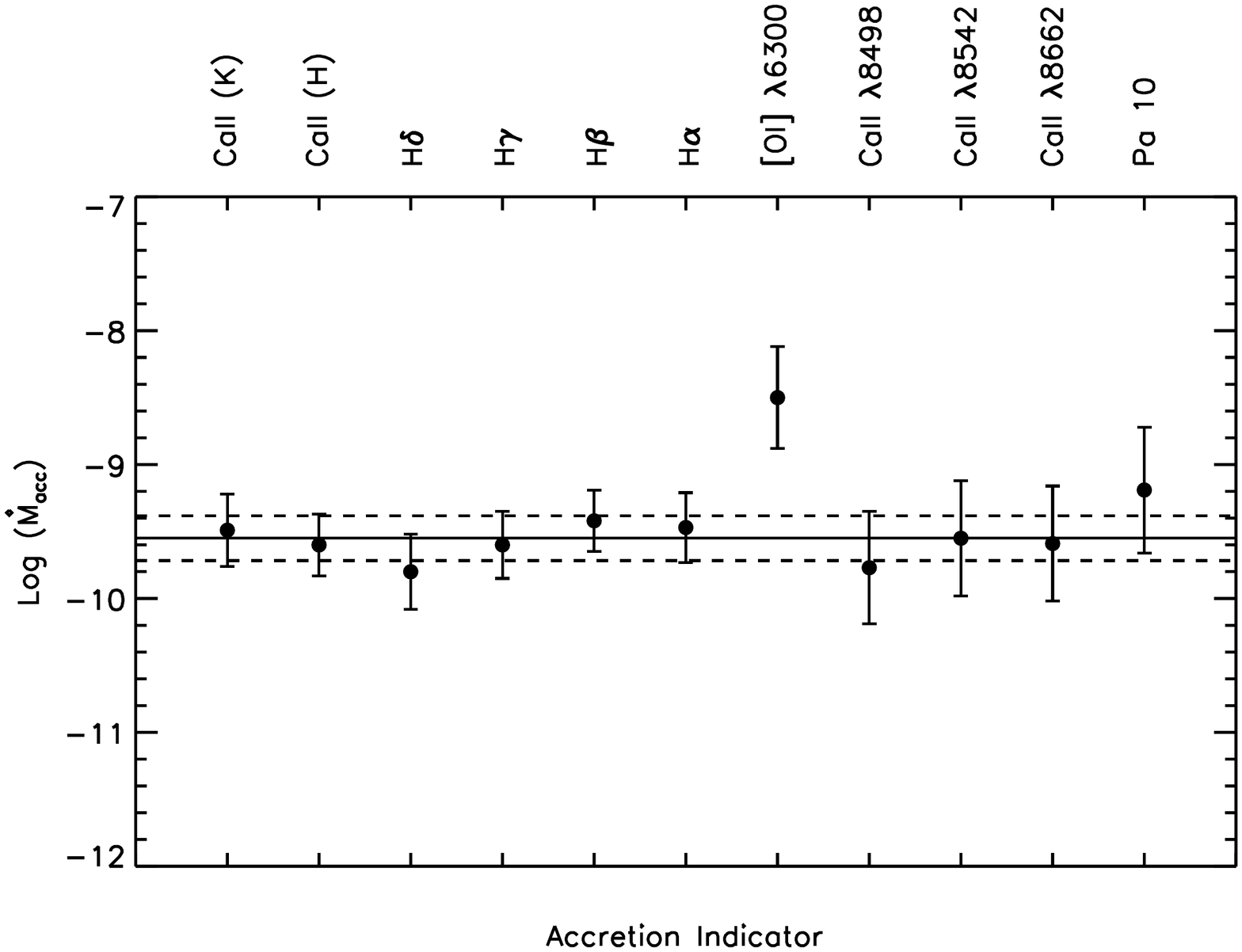}
\caption{The accretion rates $\log$ $\dot{M}_{acc}$ in $M_{\sun}$ yr$^{-1}$ derived from various line diagnostics (labelled) for HH 1158. Horizontal line indicates the mean level calculated without including the [O~{\sc i}] line. }
\label{accretion}
\end{figure}

We have estimated the mass accretion rate, $\dot{M}_{acc}$, and the mass outflow rate, $\dot{M}_{out}$, for HH 1158, by applying the same methods as used in Riaz et al. (2015). We refer to that work for further details on $\dot{M}_{acc}$ and $\dot{M}_{out}$ calculation and error estimation. Figure~\ref{accretion} shows $\dot{M}_{acc}$ derived using multiple accretion activity indicators. The value for $\dot{M}_{acc}$ derived using the [O~{\sc i}]$\lambda$6300 line luminosity is about an order of magnitude higher than the mean value. As argued in Riaz et al. (2015), considering that the [O~{\sc i}] line forms above the envelope/disk system, this indirect tracer of accretion is less obscured and could provide a more reliable measure of $\dot{M}_{acc}$ compared to other diagnostics. The mean $\dot{M}_{acc}$ for HH 1158, excluding the estimate from the [O~{\sc i}]$\lambda$6300 line, is measured to be (3.0$\pm$1.0)x10$^{-10}$ M$_{\sun}$ yr$^{-1}$. The mass outflow rate for HH 1158 derived from the [S~{\sc ii}] FELs is (5.2$\pm$2.6)x10$^{-10}$ M$_{\sun}$ yr$^{-1}$, respectively. Considering that the HH 1158 jet is also asymmetric in electron density, we estimate $\dot{M}_{out}$ of (4$\pm$2)x10$^{-10}$ M$_{\sun}$ yr$^{-1}$ in the blue component, and (1.5$\pm$0.7)x10$^{-10}$ M$_{\sun}$ yr$^{-1}$ in the red component, indicating an asymmetry in the outflow rate. The $\dot{M}_{acc}$ and $\dot{M}_{out}$ estimates presented in this work are more reliable than the estimates from Riaz et al. (2015). We have used additional accretion indicators, and have obtained spectra along the jet, which provides a better measure of $n_{e}$, the [S~{\sc ii}] fluxes, and the velocities. The high quality of the UVES spectra and a better subtraction of the nebular component also contribute to more accurate estimates. In comparison, the mass loss rates derived for HH 444, 445, and 446 are in the range of (2--8)x10$^{-7}$ M$_{\sun}$ yr$^{-1}$ (Andrews et al. 2004). Overall, the accretion and outflow rates for HH 1158 are within the range of 10$^{-7}$ -- 10$^{-10}$ M$_{\sun}$ yr$^{-1}$ observed among low-mass Class I protostars (e.g., Hillenbrand et al. 2012; Antonucci et al. 2011), and are comparatively higher than the typical rates of the order of 10$^{-10}$ -- 10$^{-12}$ M$_{\sun}$ yr$^{-1}$ observed for Class II very low-mass stars and brown dwarfs (e.g., Whelan et al. 2009; Alcal\'{a} et al. 2014). 

%A detailed study into the trends in the accretion and outflow activity rates with the evolutionary stage of the system can be conducted once a larger number of Class I very low-mass stars and brown dwarfs have been studied. 

\section{Discussion: Similarities with HH 444 -- HH 447 jets in $\sigma$ Orionis} 
\label{discuss}

A detailed study of the irradiated HH 444 -- HH 447 outflows and their driving sources was presented in Andrews et al. (2004), using high-resolution spectroscopy from the Keck HIRES spectrograph. All of these previously known $\sigma$ Ori jets demonstrate a jet/counterjet brightness asymmetry. In particular, the PV diagrams for HH 444, 445, and 446 show a bipolar asymmetric jet morphology, where the peak in emission is blue-shifted for all FELs. The blue-shifted lobe is brighter, denser, and is at a higher velocity than the red-shifted component. There is either a very weak red-shifted lobe or an extended red-shifted wing, and in all cases it is significantly fainter than the blue-shifted lobe. Several discrete knots of roughly the same velocity and intensity are also seen in the blue-shifted branches in the FEL emission for these jets. Faint, extended emission is also observed in the H$\alpha$ line for HH 445 and 446. HH 447 is the weakest among these four jets and exhibits a similar jet/counterjet structure; however, only a red-shifted branch is observed for this jet with no evidence of a blue-shifted component.

The PA for the blue lobes in these jets are 70$^{\circ}$ for HH 444, 100$^{\circ}$ for HH 445, 166$^{\circ}$ for HH 446, and 210$^{\circ}$ for HH 447 (Andrews et al. 2004). As can be visualized from Fig.~\ref{field}, the weaker outflow component for all of these HH jets is the beam oriented towards the west and is exposed to the ionizing radiation from $\sigma$ Ori, resulting in a higher extent of photo-evaporation, whereas the eastward beam facing away from $\sigma$ Ori is much brighter than the counter jet. For the case of HH 1158, the jet PA is estimated to be 102$^{\circ}$, which would imply that the brighter, blue-shifted component is oriented towards the south-east of the central source, while the weaker red component is oriented towards the north-west. HH 1158 appears to be a case more similar to HH 446 and HH 447. The orientation of these jets is such that both lobes are exposed towards $\sigma$ Ori, and thus probably receive equal amounts of radiation, unlike the more extreme cases of HH 444 and 445. Yet there are clear asymmetries seen for these jets, notably for HH 447 where the blue lobe is photoevaporated to such an extent that it was never detected. It may be the case that the brighter component is somewhat protected by the circumstellar disk from the external radiation, whereas the weaker component, which also appears to be more {\it tilted} towards $\sigma$ Ori for these three jets (Fig.~\ref{field}), has been exposed to a higher dose of photoionization from the massive stars. That the brighter jet component is probably protected to some extent by the disk has also been suggested for HH 446 (Andrews et al. 2004).

%Most irradiated outflows show asymmetries between jet and counterjet, similar to non-irradiated outflows. The asymmetries are observed in brightness, velocities, electron densities, morphologies, and even the number of knots and the mass outflow rates for the blue and red-shifted lobes. 

All of the features seen in the HH 444-447 and HH 1158 jets are also observed in non-irradiated jets. A distinction between asymmetric irradiated and non-irradiated micro-jets is more difficult to make for HH objects in H II regions, where it is easy to suspect a bright rim or compact globule to be an HH bow shock. In such cases, a criteria used for irradiated jets is to have a H$\alpha$/[S~{\sc ii}] ratio that is clearly larger than one, indicating enhanced intensity of the [S~{\sc ii}] and [N~{\sc ii}] FELs with respect to H$\alpha$ emission (e.g., Bally \& Reipurth 2003; Reipurth et al. 2010; Camer\'{o}n et al. 2013). Considering the red-shifted knot in the H$\alpha$ emission for HH 1158, the H$\alpha$/[S~{\sc ii}] ratio is 2.8. The value for this ratio is many times $>$ 1 for HH 444 -- 447 jets (Reipurth et al. 1998; Reipurth \& Bally 2001; Bally \& Reipurth 2003). While this is not an affirmative test for irradiated jets, the H$\alpha$/[S~{\sc ii}] ratio being significantly larger than the background value indicates that the blobs of material observed in an irradiated region are related to shocked material. The irradiated environment in the $\sigma$ Ori region being dominated by the UV radiation field of the massive OB stars is a strong argument to suggest that environmental effects are the main cause of the asymmetries observed for the HH 444 -- HH 447 and HH 1158 jets. 

%A few notable examples of asymmetric jets among very low-mass stars and brown dwarfs are ESO-H$\alpha$ 574, RW Aur, and ISO-ChaI 217 (Whelan et al. 2009; 2014). 

HH 1158 fits into the overall picture observed for the HH 444 -- HH 447 jets, with the evidence being that (a) it shows bipolar asymmetric jet morphology, with the lobe tilted towards $\sigma$ Ori being the weaker component. The asymmetries are also observed in the brightness, density, and velocity of the lobes, similar to HH 444-447; (b) it lies in close proximity to $\sigma$ Ori, closer than HH 447, to be ``bathed'' in the UV radiation field and termed as an irradiated jet. The high value for the H$\alpha$/[S~{\sc ii}] ratio is also consistent with the ratios measured in other irradiated jets, including HH 444 -- 447; (c) the driving source for HH 1158 is not an embedded but an optically visible source, with the jet emission seen brightly in optical lines. The driving sources for HH 444 -- HH 447 are suggested to be of mid-K to early-M spectral types (Andrews et al. 2004), corresponding to bolometric luminosities higher than $\sim$0.5 $L_{\sun}$. This makes HH 1158 the lowest luminosity analog of externally irradiated HH jets identified yet in the $\sigma$ Orionis cluster. 

A search through the literature shows that this is the lowest mass case among all known irradiated HH jets. The presence of an extended collimated jet that is bipolar and the evidence of shocked emission knots make HH 1158 a unique new HH jet at the very low-luminosity end. The comparison with HH 444 -- 447 presented here suggests that HH 1158 is a scaled down case of irradiated jets driven by YSOs. Therefore, irradiated jets can now be added to the list of properties which jets driven by the lowest luminosity objects have in common with protostellar jets.

%HH jets from very low-luminosity sources are relatively new, and we can gain a deeper insight into their characteristics as more such cases are identified. 

%A search through the literature shows that this is the lowest mass case among all known HH jets and therefore also the lowest mass source with an irradiated jet. 

\section{Summary}

We have identified a new externally irradiated jet, HH 1158, in $\sigma$ Orionis, the overall properties of which are similar to the previously known irradiated HH jets in this cluster. HH 1158 is driven by an optically visible Class Flat source, and shows bipolar asymmetric jet morphology, with the weaker jet beam tilted towards the massive stars in the cluster. There are also asymmetries observed in the brightness, density, and velocity of the lobes. Due to the close proximity to $\sigma$ Ori, the asymmetry is likely caused by the irradiated environment in the cluster. HH 1158 is the lowest luminosity analog of externally irradiated HH jets identified yet in the $\sigma$ Orionis cluster, and is the lowest luminosity case among all known irradiated HH jets.

\acknowledgements

We thank B. Reipurth, J. Bally, and J. Caballero for their suggestions on the nature of this object. BR acknowledges funding from the Marie Sklodowska-Curie Individual Fellowship (Grant Agreement No. 659383). E. T. Whelan acknowledges financial support from the Deutsche Forschungsgemeinschaft through the Research Grant Wh 172/1-1. Based on observations collected with UVES at the Very Large Telescope on Cerro Paranal (Chile), operated by the European Southern Observatory.

\onecolumn

\clearpage

\clearpage

\begin{table*}
\center
\caption{Emission line fluxes and equivalent widths}
\label{fluxes}
\begin{tabular}{ccccc}
\hline
\hline
Line & $\lambda_{central}$  & Line Flux  & Equivalent Width    \\ 
         & ($\AA$)   & (10$^{-15}$ erg cm$^{-2}$ s$^{-1}$) &  ($\AA$)  \\
\hline

H$\eta$         			& 3835.38  &  1.93$\pm$0.5 & $-$34.9$\pm$2.0    \\
H$\zeta$        			& 3889.05  & 1.74$\pm$0.5 & $-$45$\pm$10    \\
Ca~{\sc ii} K   			& 3933.66 & 2.5$\pm$0.3  & $-$20.4$\pm$4.0    \\
H$\epsilon$ + Ca~{\sc ii} H  	& 3970.07 & 1.9$\pm$0.3  & $-$7.4$\pm$0.3     \\ \relax
[S~{\sc ii}]  			& 4068.6 & 8.2$\pm$1.0  & $-$45.8$\pm$5.0     \\  \relax
[S~{\sc ii}]  			& 4076.3 & 2.5$\pm$0.5  & $-$30.0$\pm$10.0    \\
H$\delta$       			& 4101.74 & 1.5$\pm$2.5  & $-$2.5$\pm$1.5     \\ 
H$\gamma$      		& 4340.47 & 3.4$\pm$0.5  & $-$15.7$\pm$8     \\
H$\beta$            		& 4861.33 &  10.6$\pm$3  & $-$23.6$\pm$10     \\  \relax
[O~{\sc i}]           		& 6300.30 &  36.8$\pm$3.0  & $-$250$\pm$130    \\    \relax
[O~{\sc i}]          		& 6363.78 & 16.0$\pm$2.0 & $-$25.6$\pm$5.0   \\   \relax
[N~{\sc ii}]    			& 6548.0 & 1.5$\pm$0.1  & $-$3.2$\pm$0.1    \\
H$\alpha$          		& 6562.85 &  98.2$\pm$5.0 & $-$102.5$\pm$20.0    \\  \relax
[N~{\sc ii}]                      	& 6583.45 &  4.0$\pm$0.6  & $-$9.7$\pm$3.0    \\   \relax
[S~{\sc ii}]                      	& 6716.44 &  4.4$\pm$0.3 & $-$17.1$\pm$3.5    \\   \relax
[S~{\sc ii}]                      	& 6730.82 &  7.6$\pm$0.5 & $-$15.1$\pm$1.0    \\ 
O~{\sc i}  				& 7773.0   &  1.06$\pm$0.2 & $-$2.6$\pm$0.3    \\
O~{\sc i}  				& 8446.36  &  3.3$\pm$0.5 & $-$7.4$\pm$0.5    \\
Ca~{\sc ii}                    	& 8498.02  &  3.5$\pm$1.0 & $-$7.2$\pm$2.0    \\
Ca~{\sc ii}                    	& 8542.09  &   5.1$\pm$1.0 & $-$12.7$\pm$2.0   \\  \relax
[Fe~{\sc ii}]  			& 8616.95   &  0.8$\pm$0.1 & $-$1.7$\pm$0.2    \\
Ca~{\sc ii}                    	& 8662.14  &  4.7$\pm$1.0 & $-$10.4$\pm$3.0    \\   \relax
[C~{\sc i}]  			& 8727.1  &  0.4$\pm$0.3 & $-$0.7$\pm$0.3    \\
H~{\sc i} (Pa 11)  		& 8863.4  &  1.2$\pm$0.2 & $-$2.12$\pm$0.6    \\
H~{\sc i} (Pa 10)  		& 9015.6  &  0.66$\pm$0.1 & $-$1.12$\pm$0.2    \\
H~{\sc i} (Pa 9) + [Fe~{\sc ii}]    	& 9229.7  &  2.22$\pm$0.4 & $-$4.02$\pm$0.4    \\

\hline             
\end{tabular}
\end{table*}

\end{document}